\documentclass[final,5p,times,twocolumn]{elsarticle}
 \biboptions{comma,sort&compress}
\usepackage{graphicx}
\usepackage{here}

\def\beq{\begin{equation}}
\def\eeq{\end{equation}}
\def\bea{\begin{eqnarray}}
\def\eea{\end{eqnarray}}
\def\bq{\begin{quote}}
\def\eq{\end{quote}}

\def\nnb{\nonumber}
\def\ga{\left(}
\def\dr{\right)}

\def\nnb{\nonumber}
\def\la{\langle}
\def\ra{\rangle}
\def\nin{\noindent}
\def\ba{\vspace*{-0.2cm}\begin{array}}
\def\ea{\end{array}\vspace*{-0.2cm}}

\def\als{\alpha_s}

\def\gg2{ \la\alpha_s G^2 \ra}
\def\gg3{g^3f_{abc}\la G^aG^bG^c \ra}
\def\ggg4{\la\als^2G^4\ra}

\begin{document}
\begin{frontmatter}

\title{Updating $\overline{m}_{c,b}(\overline{m}_{c,b})$  from  SVZ-Moments and their Ratios}

 \author[label1]{Stephan Narison
 }
   \address[label1]{Laboratoire
Univers et Particules , CNRS-IN2P3,  
Case 070, Place Eug\`ene
Bataillon, 34095 - Montpellier Cedex 05, France.}
\ead{snarison@yahoo.fr}
\pagestyle{myheadings}
\begin{abstract}
\noindent
Using recent values of $\alpha_s$, the gluon condensates $\la\alpha_s G^2\ra$ and $\la g^3 f_{abc}G^3\ra$
and the new data on the $\psi/\Upsilon$-families,   we update our determinations of the ${\overline{MS}}$ running  quark masses $\overline{m}_{c,b}(\overline{m}_{c,b})$ from  the SVZ-Moments ${\cal M}_n(Q^2)$ and their ratios\,\cite{SNcb12,SNcb} by including higher order perturbative (PT) corrections, 
non-perturbative (NPT) terms up to dimension $d=8$ and using the degree $n$-stability criteria  of the (ratios of) moments. 
Optimal results from different (ratios of) moments converge to the accurate mean values: $\overline{m}_{c}(\overline{m}_{c})=1264(6)~{\rm MeV}$
 and  $\overline{m}_b(\overline{m}_b)=4188(8)$ MeV in Table\,\ref{tab:res}, which  improve
and confirm our previous findings\,\cite{SNcb12,SNcb} and the recent ones from Laplace sum rules\,\cite{SNcb18}. Comments on some other determinations of  $\overline{m}_{c}(\overline{m}_{c})$ and $\la\alpha_s G^2\ra$ from the SVZ-(ratios of) moments in the vector channel are given in Section 5. 
 \end{abstract}

\end{frontmatter}


\section{Introduction and SVZ-Moments}
In Refs.\,\cite{SNcb12,SNcb}, we have used  different ${\cal M}_n(Q^2)$  moments and their ratios $r^n/r^{n+j}$ introduced by SVZ\, \cite{SVZa,SVZb}\,\footnote{For reviews, see e.g.\,\cite{ZAKA,SNB1,SNB2,SNB3,SNB4,RRY,IOFFE1,RAF,YND}.} for extracting 
the values of the charm and bottom running quark masses $\overline{m}_{c,b}(\overline{m}_{c,b})$ and the dimension 4: $\la \alpha_s G^2\ra$ and 6:  $ \la g^3f_{abc} G^3\ra $ gluon condensates. Using the recent values of the gluon condensates from Laplace sum rules\,\cite{SNcb18,SNcb3} and new data on the $\psi/\Upsilon$-families masses and leptonic widths\,\cite{PDG}, we shall improve in this paper our previous results for the quark masses. 
Here, we shall be concerned with the two-point correlator: 
 \beq
\hspace*{-0.75cm} -\ga g^{\mu\nu}q^2-q^\mu q^\nu\dr \Pi_\Psi(q^2)\equiv 
 i\hspace*{-0.15cm}\int\hspace*{-0.15cm} d^4x ~e^{\rm-iqx}\la 0\vert {\cal T} J^\mu_\Psi(x)\ga J^\nu_\Psi(0)\dr^\dagger \vert 0\ra,
 \eeq
associated to the $J_\Psi^\mu=\bar \Psi \gamma^\mu \Psi$ ($\Psi\equiv c,b$) heavy quark neutral vector current. 
The corresponding moments are\,\footnote{We shall use the same normalization as  \cite{IOFFE} and some of the expressions given there.}:
 \bea
 {\cal M}_n\ga -q^2\equiv Q^2\dr&\equiv& 4\pi^2{(-1)^n\over n!}\ga {d\over dQ^2}\dr^n \Pi(-Q^2)\nnb\\
 &=&\int_{4m_Q^2}^\infty dt {{R}(t,m_c^2)\over (t+Q^2)^{n+1}}~.
 \eea
Their ratios read:
 \beq
 r_{n/n+1}(Q^2)={{\cal M}_n(Q^2)\over {\cal M}_{n+1}(Q^2)},~~~~r_{n/n+2}(Q^2)={{\cal M}_n(Q^2)\over {\cal M}_{n+2}(Q^2)}~,
 \eeq
where the experimental sides are more precise than that of the moments ${\cal M}_n(Q^2)$. It has been noticed
by \cite{NIKOLa,NIKOLb} that the OPE of ${\cal M}_n(0)$ breaks down for higher values of $n$, while it has also been mentioned in \cite{SNcb12,SNcb} that low moments $n\leq 3$ are sensitive to the way for parametrizing the high-energy part of the spectral function (hereafter called QCD continuum) making the results obtained from low moments model-dependent. Therefore, one should look for compromise values of $n$ (stability in $n$) where both problems are avoided. Another way out is to work with the
$Q^2\not=0$ moments\,\cite{RRY} where the OPE converges faster while the QCD continuum contributions are strongly suppressed. 
\section{Expressions of the SVZ-Moments ${\cal M}_n(Q^2)$}
\label{sec:qcd}
The QCD expressions of the moments can be derived from the ones of $R$. The on-shell expression of the spectral function is transformed into the $\overline{MS}$-scheme by using the known relation between the on-shell and $\overline{MS}$-scheme running quark masses. The sources of different PT contributions up to order $\alpha_s^3$ for ${\cal M}_n(Q^2 =0)$ and up to order $\alpha_s^2$ for ${\cal M}_n(Q^2 \not=0)$ are quoted in \,\cite{SNcb12} and will not be re-quoted here. The same for the different NP contributions up to dimension $d=8$ where one notice that the $d=4$ condensate contribution is known to NLO.  Some explicit numerical QCD expressions of the moments can be found in Ref. \cite{SNcb12}. We shall use the QCD parameters given in Table\,\ref{tab:param}. To the value of $\alpha_s(M_Z)$ quoted there, correspond:
\beq
\alpha_s(\overline{m}_c)=0.397(15)~~~~{\rm and}~~~~\alpha_s(\overline{m}_b)=0.227(7)~,
\eeq
where we have used the recent determinations from a recent global fit of the (axial-)vector and (pseudo)scalar charmonium and bottomium systems using Laplace sum rules\,\cite{SNcb18}: 
\beq
\overline{m}_c(\overline{m}_c)=1264(10)~{\rm MeV}~, ~~~\overline{m}_b(\overline{m}_b)=4.184(9)~{\rm MeV},
\label{eq:lsr}
\eeq
 \vspace*{-.5cm}
{\scriptsize
\begin{center}
\begin{table}[hbt]
\setlength{\tabcolsep}{0.3pc}
 \caption{\scriptsize QCD parameters }
{\small
\begin{tabular}{llll}
&\\
\hline
Dimension $d$&Name&Values [GeV$^d$]&Refs. \\
\hline
\\
0&$\alpha_s(M_Z)$&0.1182(19)&\cite{PDG,BETHKE,PICH,SALAM,SNcb18}\\
4&$\la\alpha_s G^2\ra$&$(6.35\pm 0.35)10^{-2}$& \cite{SNcb18}\\
6 &$\la g^3 f_{abc}G^3\ra$&$(8.2\pm 1.0){\rm GeV}^2\la\alpha_s G^2\ra$& \cite{SNcb3}\\
8&$\la G^4\ra$& $ (0.75\pm 025)\la G^2\ra^2$& \cite{NIKOLb,BAGAN}\\
\\
\hline
\end{tabular}
}
\label{tab:param}
\end{table}
\end{center}
}
\nin
\\

\nin 
The low-energy part of the spectral function is well described by the sum of different resonances contributions within a narrow width approximation (NWA). For the $c$-quark channel, it reads:
 \beq
\hspace*{-0.5cm} {R_c}(t)\equiv 4\pi{\rm Im} \Pi_c(t+i\epsilon)
 ={\pi N_c\over Q_c^2 \alpha^2}\sum_{J/\psi}M_{\psi}\Gamma_{\psi\to e^+e^-}\delta\ga t-M^2_{\psi}\dr,
 \eeq
 where $N_c=3$; $M_{\psi}$ and $\Gamma_{\psi\to e^+e^-}$ are the mass and leptonic width of the $J/\psi$ mesons; $Q_c=2/3$ is the charm electric charge in units of $e$; $\alpha=1/133.6$ is the running electromagnetic coupling evaluated at $M^2_{\psi}$. We shall use the experimental values of the $J/\psi$ parameters compiled in Table \ref{tab:psi}.\\
 \vspace*{-.5cm}
{\scriptsize
\begin{table}[hbt]
\begin{center}
\setlength{\tabcolsep}{1.5pc}
 \caption{\scriptsize    Masses and electronic widths of the  $J/\psi$ family from PDG 16 \cite{PDG}. }
{\small
\begin{tabular}{lll}
&\\
\hline
Name&Mass [MeV]&$\Gamma_{J/\psi\to e^+e^-}$ [keV] \\
\hline
\\
$J/\psi(1S)$&3096.916(11)&5.55(14)\\
$\psi(2S)$&3686.097(25)&2.34(4)\\
$\psi(3770)$&3773.13(0.35)&0.262(18)\\
$\psi(4040)$&4039(1)&0.86(7)\\
$\psi(4160)$&4191(5)&0.48(22)\\
$\psi(4415)$&4421(4)&0.58(7)\\
\\
\hline
\end{tabular}
}
\end{center}
\label{tab:psi}
\end{table}
}
\nin
\\
We shall parametrize the contributions from $\sqrt{t_c}\geq (4.5\pm 0.1)$ GeV using either:

-- {\it Model 1:} The approximate PT QCD expression of  the spectral function to order $\alpha_s^2$ up to order $(m_c^2/t)^6$ given in \cite{CHET3} and the $\alpha_s^3$ contribution from non-singlet contribution up to order $(m_c^2/t)^2$ given in \cite{HOANG}. 

-- {\it Model 2:} The asymptotic PT expression of the  spectral function known  to order $\alpha_s^3$ where the quark mass corrections are neglected~\footnote{Original papers are given in Refs. 317 to 321 of the book in Ref.~\cite{SNB1}.}.

-- {\it Model 3:} Fits of different data above the $\psi(2S)$ mass: we shall take e.g the results in \cite{HOANG} where a comparison of results from different fitting procedures can be found in this paper (see e.g\, \cite{KUHN1}). 
\section{Running $\overline{m}_c(\overline{m}_c)$ charm quark mass from ${\cal M}_n(0)$}
\begin{figure}[hbt]
\begin{center}
\includegraphics[width=9cm]{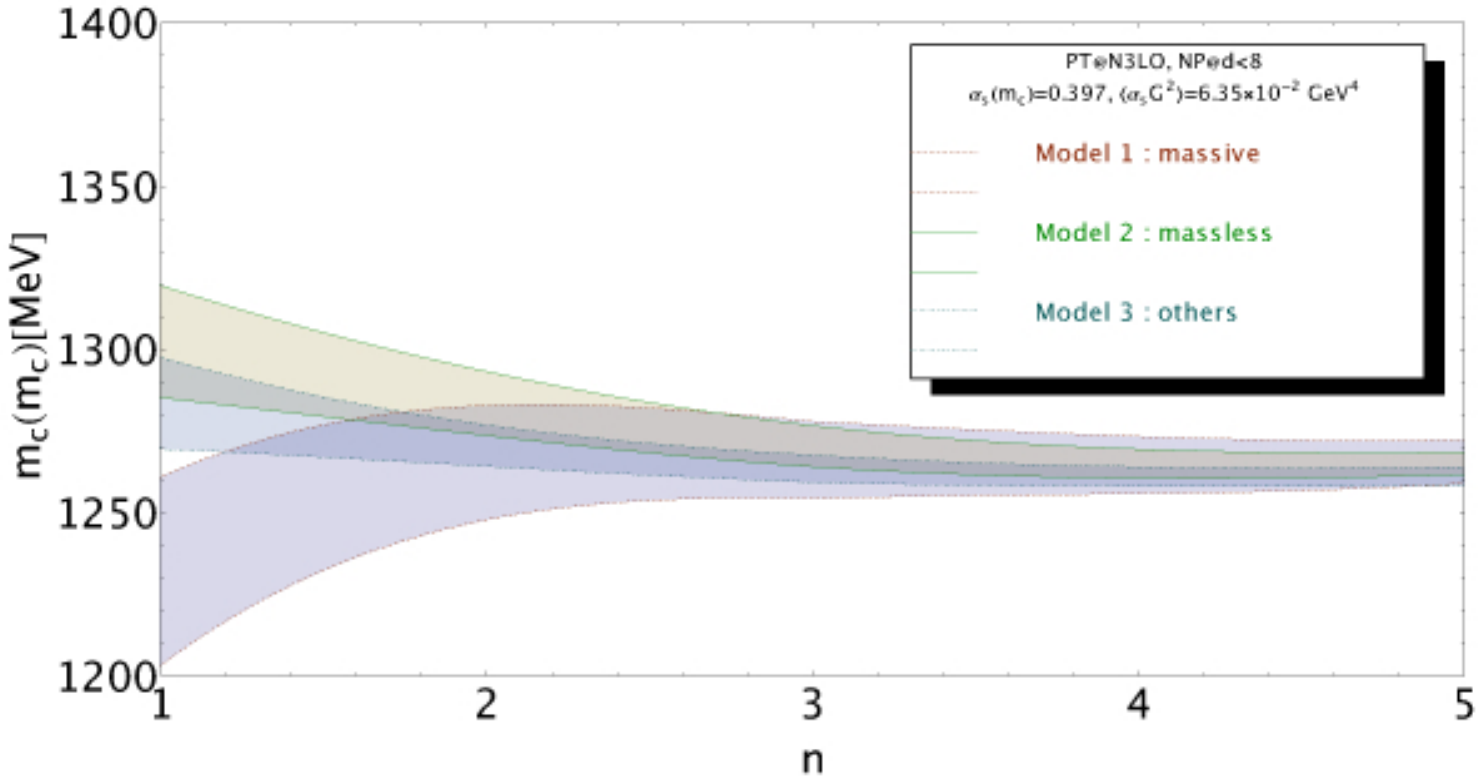}
\vspace*{-0.25cm}
\caption{\footnotesize  Values of $\overline{m}_c(\overline{m}_c)$ from from ${\cal M}_n(0)$ for different values of $n$ using the QCD input parameters in Table\,\ref{tab:param} and the three models given previously for the QCD continuum parametrization.} 
\label{fig:mom-c}
\end{center}
\vspace*{-0.5cm}
\end{figure}

-- Using the previous models for parametrizing the QCD continuum, we show in Fig.\ref{fig:mom-c} the values of 
$\overline{m}_c(\overline{m}_c)$ from ${\cal M}_n(0)$ for different values of $n$. We have used the Mathematica program Find Root for extracting the values of $\overline{m}_c(\overline{m}_c)$ left as a free parameter in the OPE including $1/\overline{m}_c^8$ corrections. 

-- One can see that the model-dependence of the results disappear for $n\geq 3$ where stability in $n$ is obtained. Noting that Model 1 gives the most conservative result and appears (a priori) to be a good approximation of the spectral function as it includes higher order radiative $\oplus$ mass corrections, we shall only consider Model 1 in the rest of the paper. At the stability point $n\simeq 3-4$, we deduce the optimal estimate  (in units of MeV):
\beq
\hspace*{-0.5cm}\overline{m}_c(\overline{m}_c)\vert^4_0=1266(8.8)_{ex}(0.7)_{\alpha_s}(5.2)_{\alpha^4_s}
(0.1)_{G^2}(0.3)_{G^3}(1.5)_{G^4}.
\label{eq:mass_0a}
\eeq

\begin{figure}[hbt]
\begin{center}
\includegraphics[width=9.5cm]{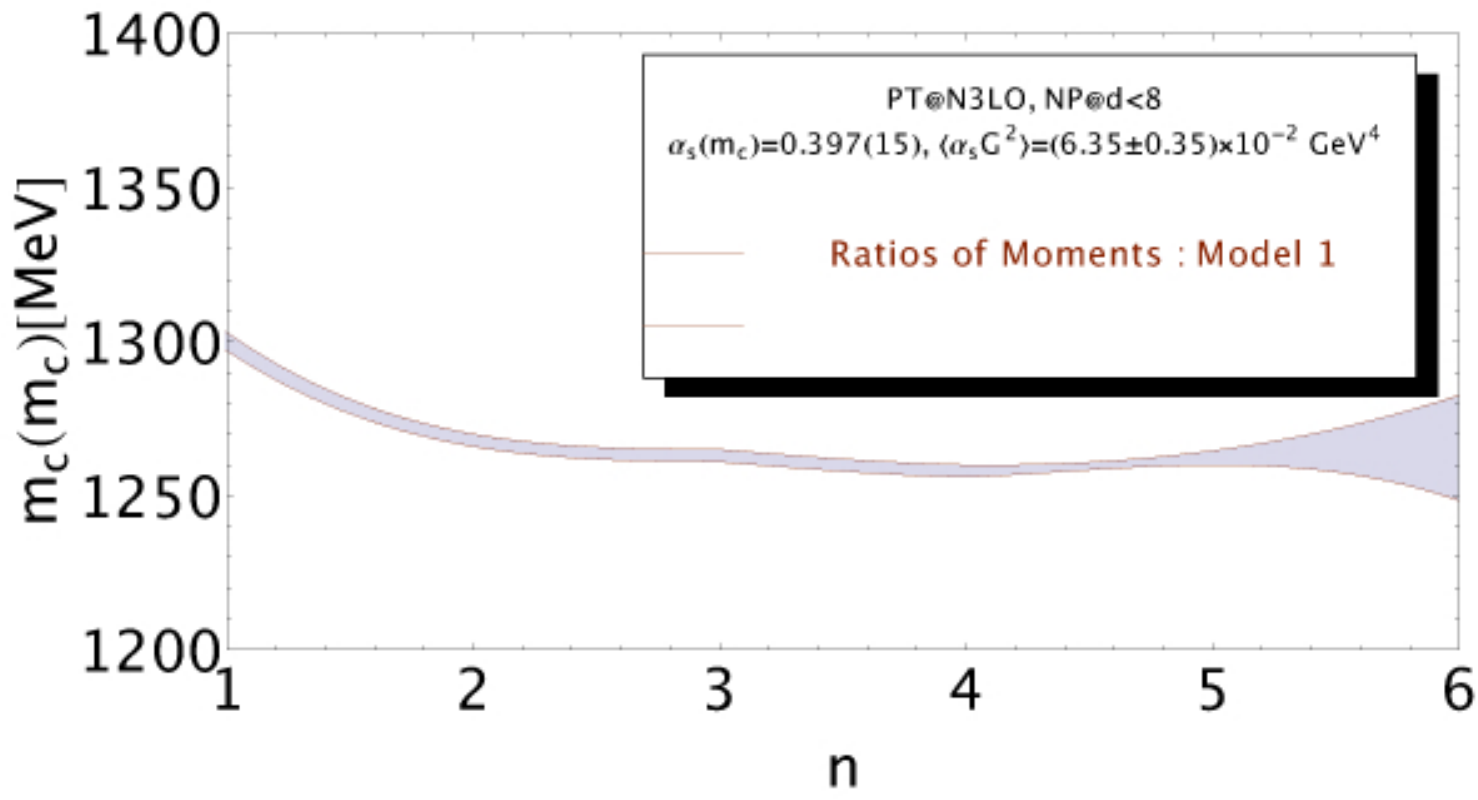}
\vspace*{-0.25cm}
\caption{\footnotesize  Values of $\overline{m}_c(\overline{m}_c)$ from the ratios of moments $r_{n/n+1}(0)$ and $r_{n/n+2}(0)$ for different values of $n$ using the QCD input parameters in Table\,\ref{tab:param} and Model 1 given previously for the QCD continuum parametrization. In the $n$ axis: $1\equiv r_{1/2}, 2\equiv r_{2/3}, 3\equiv r_{2/4}, 4\equiv r_{3/4}, 5\equiv r_{3/5}, 6\equiv r_{4/5}$} 
\label{fig:ratio-c}
\end{center}
\end{figure}

-- We do a similar analysis for the ratios of moments $r_{n/n+1}(0)$ and $r_{n/n+2}(0)$. The results versus the degree of moments are shown in Fig.\,\ref{fig:ratio-c}. We deduce, at the stability point $n\simeq 4$, the value  (in units of MeV):
\beq
\hspace*{-0.5cm}
\overline{m}_c(\overline{m}_c)\vert^{3/4}_0=1264(0.1)_{ex}(2.7)_{\alpha_s}(9.9)_{\alpha^4_s}
(0.3)_{G^2}(0.2)_{G^3}(4.3)_{G^4},
\label{eq:mass_0b}
\eeq
where one can notice that the experimental error is reduced compared to the moment results while the ones induced by the QCD parameters have increased. 

-- The errors from the $\alpha_s^4$-term is assumed to be about 
the size of the contribution from the known $\alpha_s^3$ term which is a generous error. 

\nin
\section{Running $\overline{m}_c(\overline{m}_c)$ charm quark mass from ${\cal M}_n(Q^2\not=0)$}
\begin{figure}[hbt]
\begin{center}
\includegraphics[width=9cm]{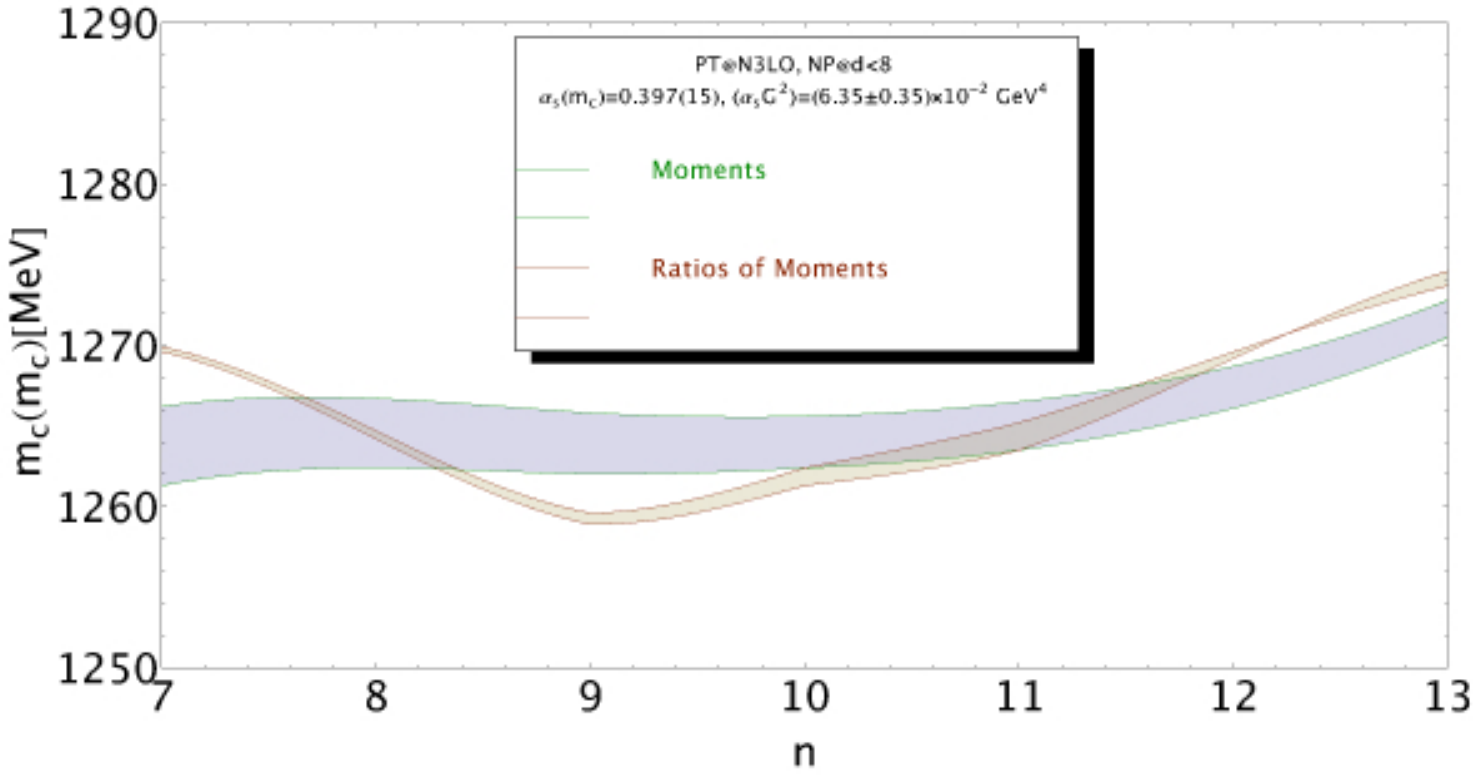}
\vspace*{-0.25cm}
\caption{\footnotesize  Values of $\overline{m}_c(\overline{m}_c)$ from the moments ${\cal M}_n(4\overline{m}_c^2)$  and their ratios $r_{n/n+1}(4\overline{m}_c^2)$ and $r_{n/n+2}(4\overline{m}_c^2)$ for different values of $n$ using the QCD input parameters in Table\,\ref{tab:param} and Model 1 given previously for the QCD continuum parametrization. In the $n$ axis: $7\equiv r_{7/8}, 8\equiv r_{7/9}, 9\equiv r_{8/9}, 10\equiv r_{8/10}, 11\equiv r_{9/10}, , 12\equiv r_{9/11}, , 13\equiv r_{10/11}$.} 
\label{fig:4mc2}
\end{center}
\end{figure}

\begin{figure}[hbt]
\begin{center}
\includegraphics[width=9cm]{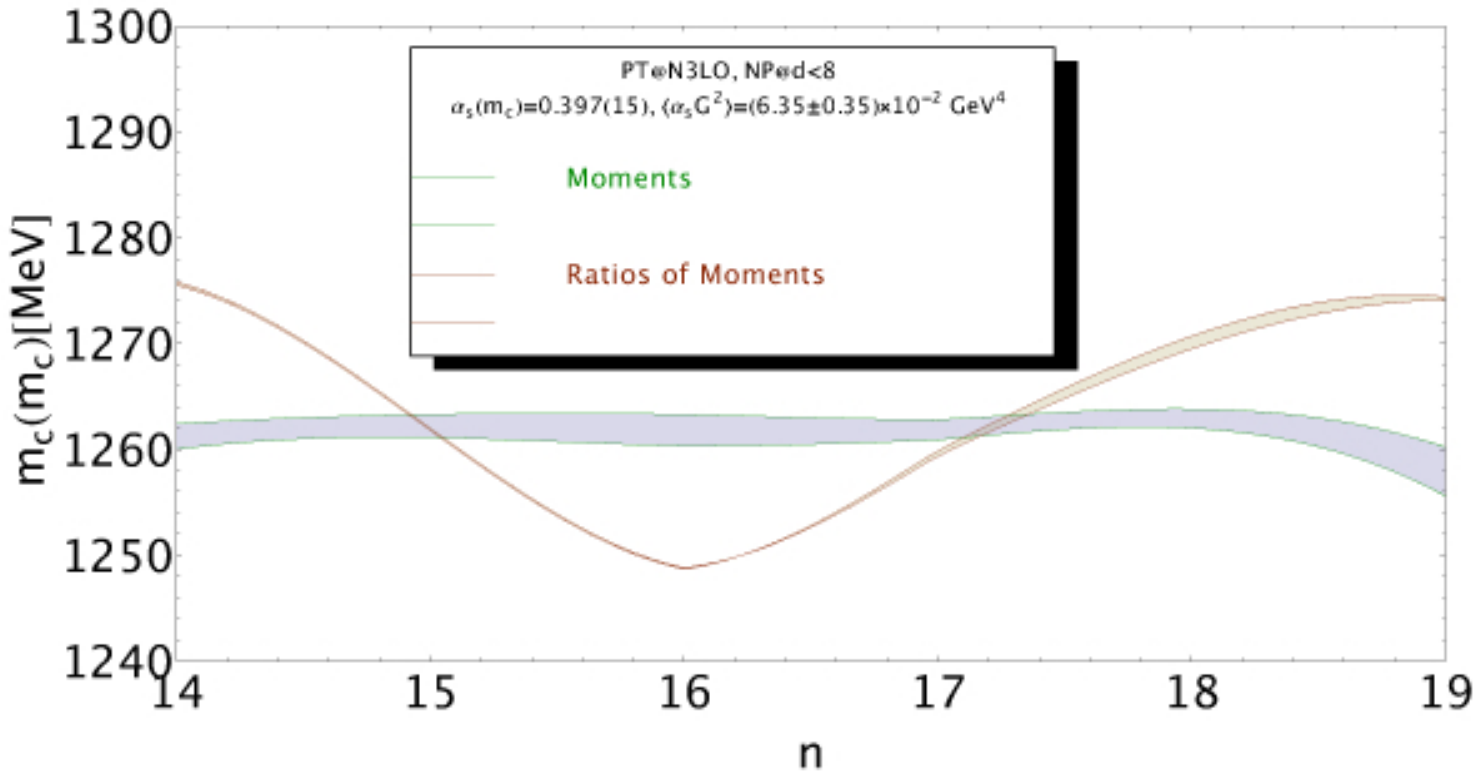}
\vspace*{-0.25cm}
\caption{\footnotesize  Values of $\overline{m}_c(\overline{m}_c)$ from the moments ${\cal M}_n(8\overline{m}_c^2)$  and their ratios $r_{n/n+1}(8\overline{m}_c^2)$ and $r_{n/n+2}(8\overline{m}_c^2)$ for different values of $n$ using the QCD input parameters in Table\,\ref{tab:param} and Model 1 given previously for the QCD continuum parametrization. In the $n$ axis: $14\equiv r_{14/16}, 15\equiv r_{15/16}, 16\equiv r_{15/17}, 17\equiv r_{16/17}, 18\equiv r_{16/18}, 19\equiv r_{17/18}$.} 
\label{fig:8mc2}
\end{center}
\end{figure}

Previous analysis can be extended to the case of $Q^2\not=0$ moments where a better convergence of 
the OPE is expected\,\cite{RRY} and where the QCD continuum contribution to the moments is smaller as we shall work with higher moments at which the $n$-stability is reached. The PT expression is known here up to order $\alpha_s^2$. 
We show the results from the (ratios of) moments in Figs.\,\ref{fig:4mc2} and \ref{fig:8mc2} for ${\cal M}_n(Q^2=4\overline{m}_c^2)$ and ${\cal M}_n(Q^2=8\overline{m}_c^2)$. 
We conclude that the most stable results come from the moments from which we deduce to order $\alpha_s^2$  (in units of MeV):
\bea
\hspace*{-1cm}\overline{m}_c(\overline{m}_c)\vert^{10}_{4\overline{m}_c^2}&\hspace*{-0.2cm}=\hspace*{-0.2cm}&1263(1.6)_{ex}(0.3)_{\alpha_s}(1.3)_{\alpha^3_s}
(0.2)_{G^2}(0.3)_{G^3}(1)_{G^4}~,\nnb\\
\hspace*{-0.5cm}\overline{m}_c(\overline{m}_c)\vert^{16}_{8\overline{m}_c^2}&\hspace*{-0.2cm}=\hspace*{-0.2cm}&1261(1)_{ex}(0)_{\alpha_s}(0.3)_{\alpha^3_s}
(0.1)_{G^2}(0.1)_{G^3}(0.8)_{G^4}~.
\label{eq:mass_8mc2b}
\eea
The previous results are collected in Table\,\ref{tab:res}. 
\begin{figure}[hbt]
\begin{center}
\includegraphics[width=9cm]{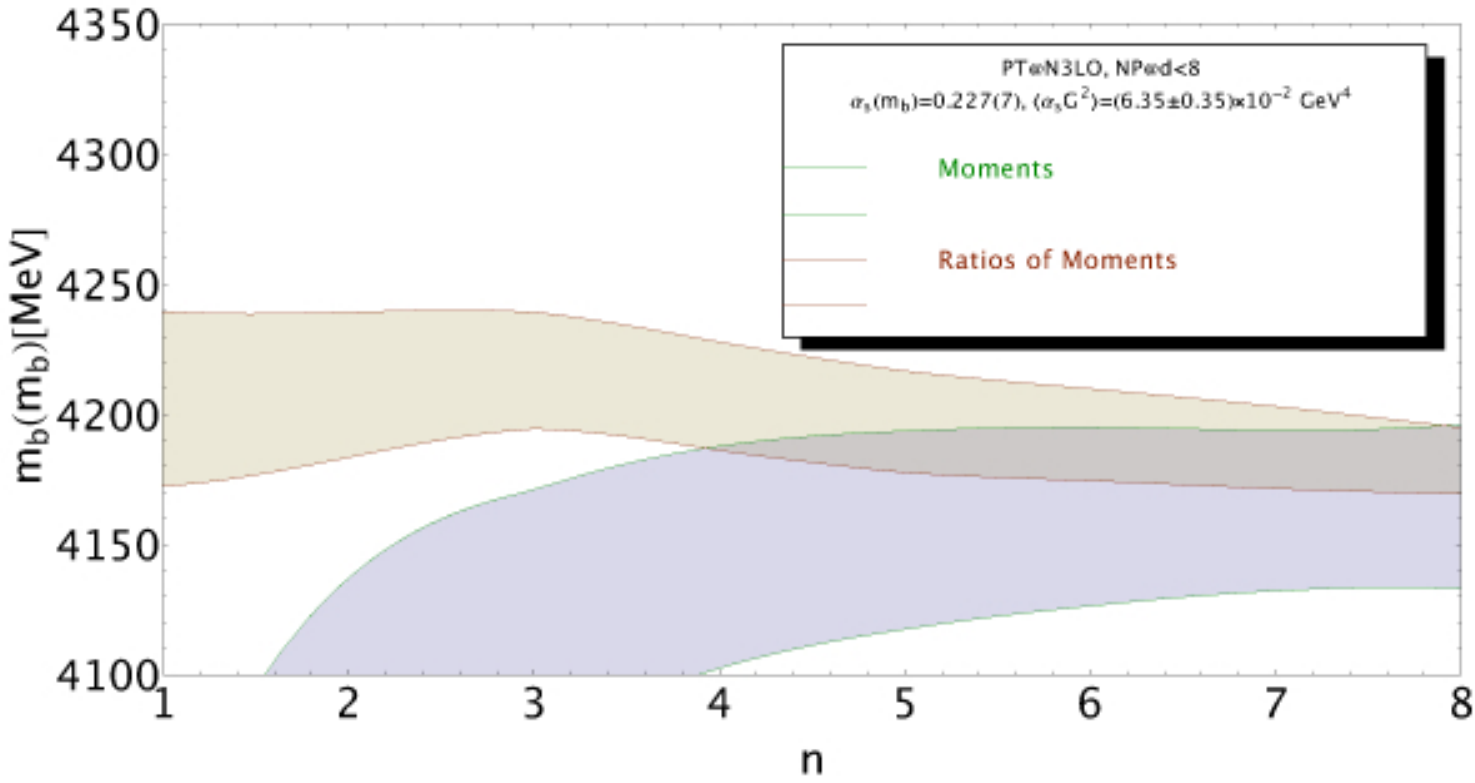}
\vspace*{-0.25cm}
\caption{\footnotesize  Values of $\overline{m}_b(\overline{m}_b)$ from the moments ${\cal M}_n(0)$  and their ratios $r_{n/n+1}(0)$ and $r_{n/n+2}(0)$ for different values of $n$ using the QCD input parameters in Table\,\ref{tab:param} and Model 1 given previously for the QCD continuum parametrization. 
In the $n$ axis: $1\equiv r_{1/2}, 2\equiv r_{2/3}, 3\equiv r_{2/4}, 4\equiv r_{3/4}, 5\equiv r_{3/5}, 6\equiv r_{4/5},
7\equiv r_{4/6}, 8\equiv r_{5/6}$.} 
\label{fig:momb0}
\end{center}
\end{figure} 
\begin{figure}[hbt]
\begin{center}
\includegraphics[width=9cm]{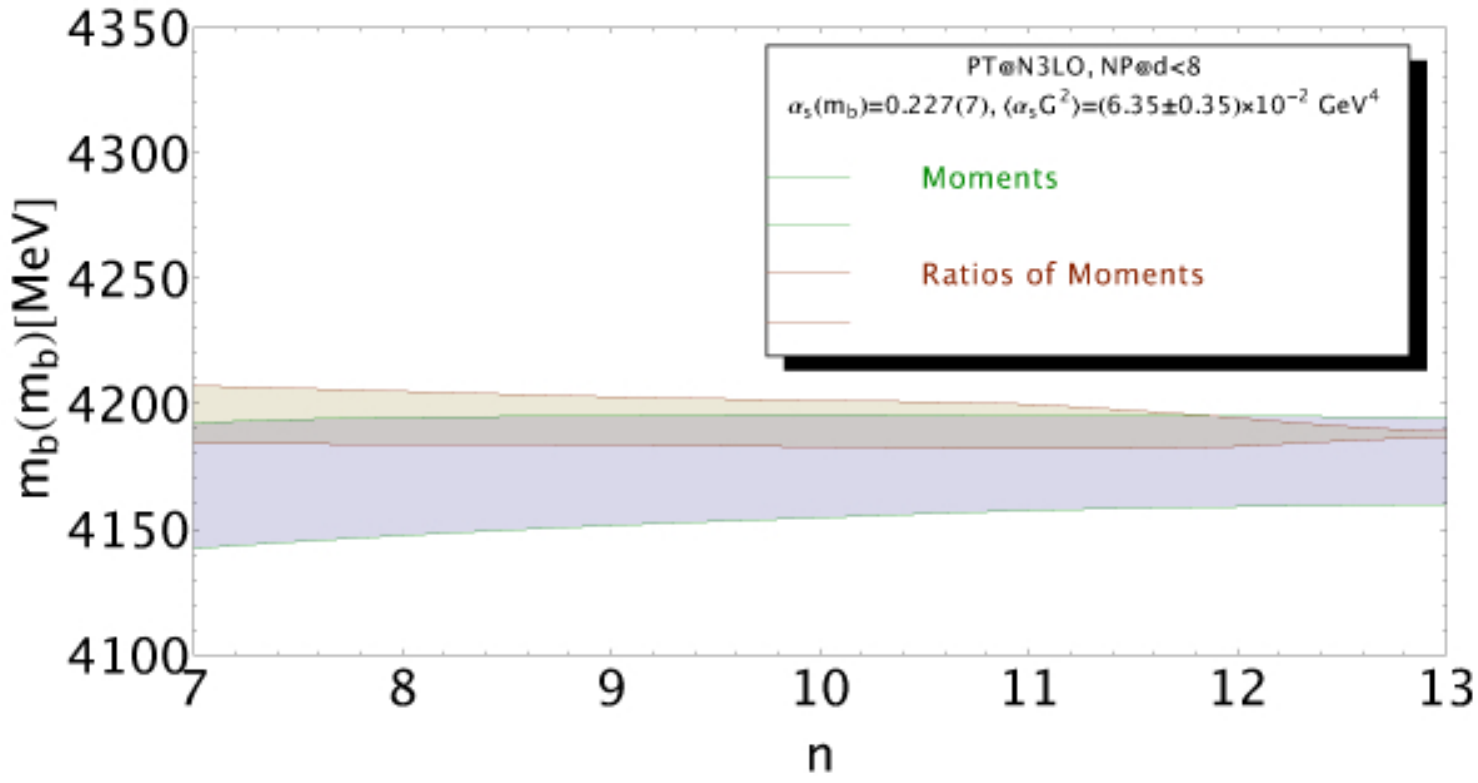}
\vspace*{-0.25cm}
\caption{\footnotesize  Values of $\overline{m}_b(\overline{m}_b)$ from the moments ${\cal M}_n(4\overline{m}_b^2)$  and their ratios $r_{n/n+1}(4\overline{m}_b^2)$ and $r_{n/n+2}(4\overline{m}_b^2)$ for different values of $n$ using the QCD input parameters in Table\,\ref{tab:param} and Model 1 given previously for the QCD continuum parametrization. In the $n$ axis: $7\equiv r_{7/8}, 8\equiv r_{7/9}, 8\equiv r_{8/9}, 9\equiv r_{8/10}, 10\equiv r_{9/10}, 11\equiv r_{9/11}, 12\equiv r_{10/11}, 13\equiv r_{10/12}$.} 
\label{fig:4mb2}
\end{center}
\end{figure} 
\begin{figure}[hbt]
\begin{center}
\includegraphics[width=9cm]{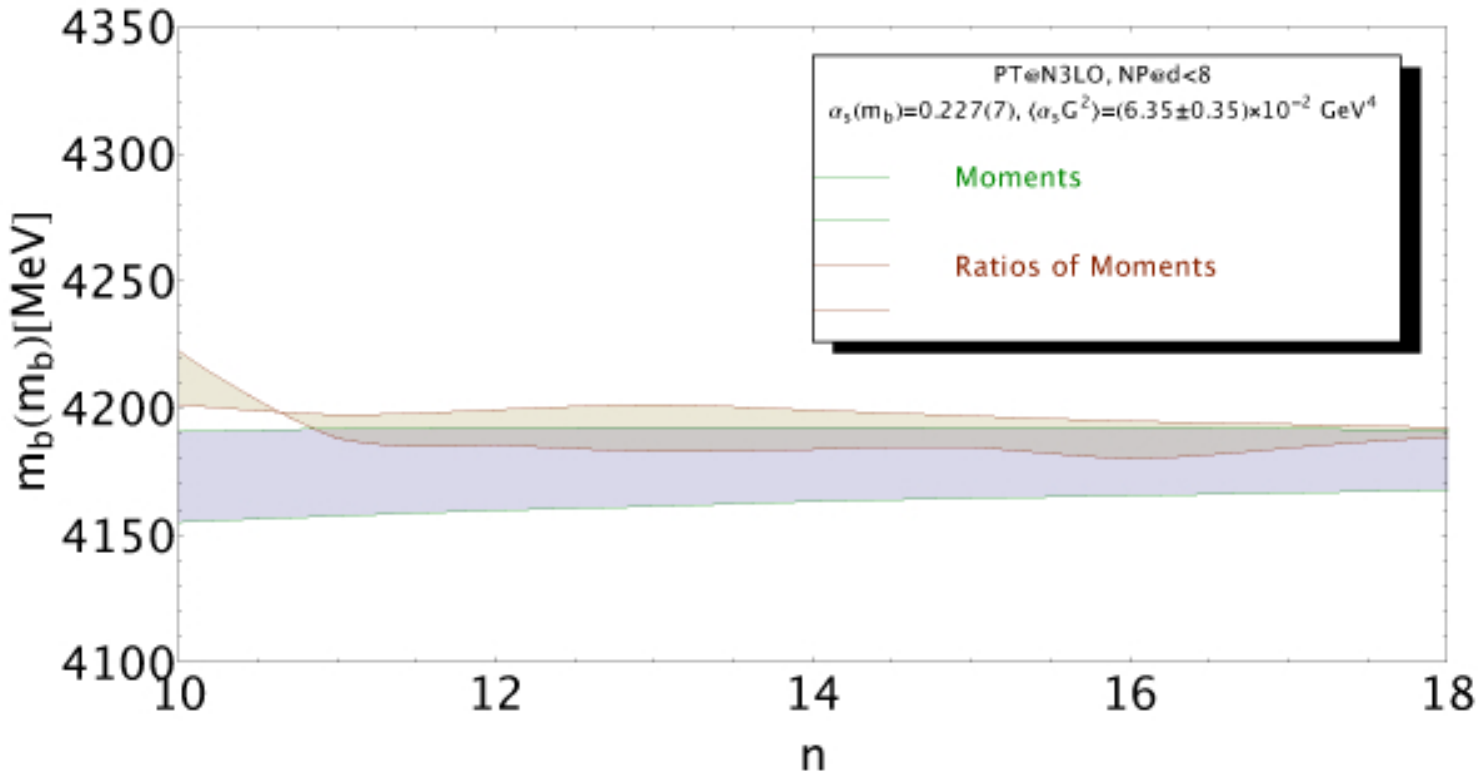}
\vspace*{-0.25cm}
\caption{\footnotesize  Values of $\overline{m}_b(\overline{m}_b)$ from the moments ${\cal M}_n(8\overline{m}_b^2)$  and their ratios $r_{n/n+1}(8\overline{m}_b^2)$ and $r_{n/n+2}(4\overline{m}_b^2)$ for different values of $n$ using the QCD input parameters in Table\,\ref{tab:param} and Model 1 given previously for the QCD continuum parametrization. In the $n$ axis: $10\equiv r_{8/10}, 11\equiv r_{9/10}, 12\equiv r_{9/11}, 13\equiv r_{10/11}, 14\equiv r_{10/12}15\equiv r_{11/12}, 16\equiv r_{12/13}, 16\equiv r_{15/17}, 17\equiv r_{12/14}, 18\equiv r_{13/14}$.} 
\label{fig:8mb2}
\end{center}
\end{figure} 
{\scriptsize
\begin{table}[hbt]
\setlength{\tabcolsep}{1.5pc}
 \caption{\scriptsize    Masses and electronic widths of the  $\Upsilon$ family from PDG 16\cite{PDG}. }
{\small
\begin{tabular}{lll}
&\\
\hline
Name&Mass [MeV]&$\Gamma_{\Upsilon\to e^+e^-}$ [keV] \\
\hline
\\
$\Upsilon(1S)$&9460.30(26)&1.340(18)\\
$\Upsilon(2S)$&10023.26(31)&0.612(11)\\
$\Upsilon(3S)$&10355.2(5)&0.443(8)\\
$\Upsilon(4S)$&10579.4(1.2)&0.272(29)\\
$\Upsilon(10860)$&10891(4)&0.31(7)\\
$\Upsilon(11020)$&$10987(^{+11}_{-3.4})$&0.13(3)\\
\\
\hline
\end{tabular}
}
\label{tab:upsilon}
\end{table}
}
\section{Comments on $\overline{m}_c(\overline{m}_c)$ and $\la\alpha_sG^2\ra$ from ${\cal M}_n(Q^2)$}
-- One can notice in Fig.\,\ref{fig:mom-c} that the values of $\overline{m}_c(\overline{m}_c)$ from the moments ${\cal M}_{n\leq 2}(0)$ are strongly affected by the QCD continuum parametrization though agree within the errors with the ones in  \cite{KUHN1,HOANG,MAIER,BOUG}. For the case of $n=1$ moment used by previous authors to extract their final results, one can deduce from Fig.\ref{fig:mom-c}:
\beq
\overline{m}_c(\overline{m}_c)\vert^1_0=1262(59)~{\rm MeV}~,
\eeq
where the error is dominated by the different parametrizations of the QCD continuum models. One can compare this result with the one $\overline{m}_c(\overline{m}_c)=1275(23)~{\rm MeV}$ from \cite{HOANG} and the improved recent estimate 1279(10) MeV from \cite{KUHN1} obtained for $\alpha_s(M_Z)=0.118$ from the analogous $n=1$ moment. The sensitivity of the results on the high energy part of the spectral function may question the accuracy of the results quoted  in these papers from ${\cal M}_{1}(0)$.  

-- Instead, in the $n-$stability region, the QCD continuum-model-dependence of the result disappears (see Fig.\,\ref{fig:mom-c}) and leads to the optimal and more accurate value given in Eq.\,\ref{eq:mass_0a}:
\beq
\overline{m}_c(\overline{m}_c)\vert^4_0=1266(9)~{\rm MeV}~.
\eeq
 The error due to the parametrization  of the spectral function is even reduced when working  with the ratio of moments (see Fig\,\ref{fig:ratio-c}) leading to the result in Eq.\,\ref{eq:mass_0b}:
 \beq
\overline{m}_c(\overline{m}_c)\vert^{3/4}_0=1264(11)~{\rm MeV}~,
\eeq
but the errors due to the QCD parameters have increased compared to the one of the moment.

-- One can also notice from the Tables in Ref.\,\cite{KUHN1} that the stability of the central values is reached from ${\cal M}_{n= 2,3}(0)$ which is about 10 MeV below their favoured choice from  ${\cal M}_{1}(0)$.  A such value is in a better agreement with our previous results quoted in Eq.\,\ref{eq:mass_0a}. 

-- We estimate the errors in the truncation of the PT series by including the $\alpha_s^4$ contribution
assumed to be of the same size as the $\alpha_s^3$ one (a geometric growth of the coefficient observed for massless quarks\,\cite{SNZ} may not be extrapolated for heavy quarks). The induced error is about 5 MeV which is smaller than the one of 19 MeV quoted in Ref.\cite{HOANG} estimated using some iterative or contour improved procedures where the effect of the subtraction scale $\mu$ is also included.

-- However, it is not clear that moving the subtraction scale from $\overline{m}_c(\overline{m}_c)$ to higher values, say 3 GeV\,\cite{KUHN1,HOANG,MAIER,BOUG} for improving the convergence of the PT series can help due to the ambiguity of the charm quark mass definitions used in the OPE [$(1/m_c)$ expansion].  Indeed apart the Wilson coefficient of $\la \alpha_s G^2\ra$ known to NLO\,\cite{BROAD}, the ones of the high-dimension condensates are only known to LO.
Refs\,\cite{KUHN1,MAIER,BOUG} choose to work with the pole mass  in the OPE which, as emphasized in\,\cite{HOANG} is ambiguous  due to the IR renormalon contribution. Then, the use of the running mass in the OPE  can be better justified which is also consistent with the use of the running mass in the PT contributions. However,  if one moves the subtraction scale $\mu$ from $\overline{m}_c(\mu)=1.264$ to 3 GeV, $\overline{m}_c(\mu)$ moves from 1.264 to  0.972 GeV which can induce an enhancement of about $1.3^d$ for the dimension $d$ condensate contributions to the moments.  Therefore, a careful analysis including radiative corrections to the Wilson coefficients of each condensate should be done when working at high values of $\mu$. 
To my knowledge, this point has not yet been carefully studied. In order to circumvent a such large enhancement, which does not arise when working with the Laplace sum rule\,\cite{SNcb18} where an optimal value of $\mu$ has been derived,  we limit here to the (usual and natural)  choice  $\mu=\overline{m}_c$ and do not try to move it arbitrarily around this value. 

-- Coulombic corrections have been roughly estimated in Ref.\,\cite{SNcb12}. However, it has been also argued in Ref. \cite{IOFFE} that this contribution, which is not under a good control, can be safely neglected in the relativistic sum rules.  Therefore, we shall not consider such corrections in this paper. 

-- In\,\cite{IOFFE,IOFFE1}, the set of QCD parameters :
\beq
\hspace*{-0.5cm}\overline{m}_c(\overline{m}_c)=1275(15)~{\rm MeV}~,~~0.7\leq\la\alpha_sG^2\ra\times 10^2\leq 6.3~{\rm GeV}^4,
\label{eq:ioffe}
\eeq
obtained from the moments used here has been favoured. Examining Figs. 4  and 5 of \,\cite{IOFFE}, one can see that the values of $\overline{m}_c(\overline{m}_c)$ from the different moments alone cannot fix accurately the values of $\la\alpha_sG^2\ra$ due to the absence of $\overline{m}_c(\overline{m}_c)$ stability versus $\la\alpha_sG^2\ra$. 
This feature has been also observed from the analysis of the same vector charmonium using Laplace sum rules\,\cite{SNcb18} where constraints from some other charmonium channels are needed for reaching more accurate results. 

-- To the value of $\la\alpha_sG^2\ra$ given  in Table\,\ref{tab:param} which is in the upper end of the range in Eq.\,\ref{eq:ioffe}, one can extract from Figs. 4 and 5 of  \,\cite{IOFFE} the value:
\beq
\overline{m}_c(\overline{m}_c)\approx 1260~{\rm MeV}~,
\eeq
which is consistent within the errors with our previous results in Table\,\ref{tab:res}. 

-- The authors deduce their favorite result in Eq.\,\ref{eq:ioffe} from a common solution of the moments and of their ratios, where one can notice, from our Figs.\,\ref{fig:4mc2} and \ref{fig:8mc2}, that, at  a fixed value of $\la\alpha_sG^2\ra$, the value of $\overline{m}_c(\overline{m}_c)$ from the ratios of moments meets the moments outside the $n$-stability of ${\cal M}_n(Q^2)$, while the ratios increase rapidly with $n$. This fact indicates that a such requirement may not be reliable. 

-- Beyond the OPE, we can also have some contributions
due to the so-called Duality Violation, which is model-dependent. It can be parametrized (within our normalization) as\,\cite{BOITO,PICH2}:
\beq
{\cal R}_c^{DV}=(4\pi^2)t^{\lambda_v}e^{-(\delta_v+\gamma_vt)}{\rm sin}(\alpha_v+\beta_vt)~,
\eeq
where the coefficients are free parameters and come from a fitting procedure. For an approximate estimate of this additional effect, we compare its contribution with the QCD continuum one  parametrized by the asymptotic expression of PT spectral function ($m_c=0$) (Model 2) from the threshold $\sqrt{t_c}$=
4.5 GeV.  We use the coefficients:
\beq
\lambda_v=0,~~\delta_v\approx 3.6,~~\gamma_v\approx 0.6,~~\alpha_v\approx -2.3,~~\beta_v\approx 4.3~,
\eeq
fixed from $\tau$-decay data by assuming that they can be applied here. We found that, in the example $n=1$ and $Q^2=0$,  this effect is completely negligible even allowing a low value of  $\sqrt{t_c}$=
1.65 GeV at which the fit of the coefficients has been performed.
\section{Running $\overline{m}_b(\overline{m}_b)$ bottom quark mass from ${\cal M}_n(Q^2)$}
The previous analysis is extended to the $b$-quark mass. We shall use the data input in Table\,\ref{tab:upsilon}.
Behaviours of the (ratios of) moments versus the degree of the moments are given in Figs.\,\ref{fig:momb0} to \ref{fig:8mb2}. We deduce as optimal values the overlapping regions of the one from the moments and the ratios of moments. We obtain to order $\alpha_s^3$ (in units of MeV):
\beq
\hspace*{-0.7cm}\overline{m}_b(\overline{m}_b)\vert^6_0=4185.9(8.2)_{ex}(4)_{\alpha_s}(1.7)_{\alpha^4_s}
(0.8)_{G^2}(0.2)_{G^3}(0.2)_{G^4}.
\label{eq:massb_0a}
\eeq
and to order $\alpha_s^2$  (in units of MeV):
\bea
\hspace*{-0.7cm}\overline{m}_b(\overline{m}_b)\vert^{10}_{4m_b^2}&\hspace*{-0.3cm}=\hspace*{-0.3cm}&4189.2(6.4)_{ex}(1.6)_{\alpha_s}(3.6)_{\alpha^3_s}
(0.5)_{G^2}(0)_{G^3}(0)_{G^4},\nnb\\
\hspace*{-0.7cm}\overline{m}_b(\overline{m}_b)\vert^{13}_{8m_b^2}&\hspace*{-0.3cm}=\hspace*{-0.3cm}&4187.7(4.3)_{ex}(1)_{\alpha_s}(5.0)_{\alpha^3_s}
(0.3)_{G^2}(0.3)_{G^3}(0.3)_{G^4}.
\label{eq:mass_8mb2b}
\eea
 These results are quoted in Table\,\ref{tab:res}. 

{\scriptsize
\begin{table}[hbt]
\begin{center}
\setlength{\tabcolsep}{2.5pc}
 \caption{\scriptsize    Charm and bottom running masses $\overline{m}_{c,b}(\overline{m}_{c,b})$ from (ratios of) moments. }
 \label{tab:res}
{\small
\begin{tabular}{ll}
&\\
\hline
Observables&Mass [MeV]\\
\hline
\\
{Charm} \\
${\cal M}^4(0)$&1266(9.0)\\
$r^{3/4}(0)$&1264(11.1)\\
${\cal M}^{10}(4\overline{m}_c^2)$&1263(2.3)\\
${\cal M}^{16}(8\overline{m}_c^2)$&1261(1.3)\\
{\it  Mean }& {\it 1264(6)}\\
\\
{ Bottom} \\
${\cal M}^6(0)\oplus r^{4/5}(0)$&4186(9.3)\\
${\cal M}^{10}(4\overline{m}_b^2)\oplus r^{9/10}(4\overline{m}_b^2)$&4189(7.5)\\
${\cal M}^{13}(8\overline{m}_b^2)\oplus r^{10/11}(8\overline{m}_b^2)$&4188(6.7)\\
{\it  Mean} & {\it 4188(8)} \\
\\
\hline
\end{tabular}
}
\end{center}
\end{table}
}
\section{Conclusions}
We have updated our previous results in Refs.\,\cite{SNcb12,SNcb} from SVZ-(ratios of) moments. These results are confirmed and improved by the new ones summarized in Table\,\ref{tab:res}. The simultaneous use of the higher moments and their ratios  reduce notably the errors in the mass determinations. Though it is difficult to estimate the systematic errors of the approach, we can expect that they are at most equal to the ones quoted in this paper. These new results are also in perfect agreement with the ones quoted in Eq.\,\ref{eq:lsr} from a recent global fit of the (axial-)vector and (pseudo)scalar charmonium and bottomium systems using Laplace sum rules\,\cite{SNcb18}. Some comments on the existing estimates of the quark masses and gluon condensates from SVZ-(ratios of) moments are given in Section 5.
Our results are comparable with recent results from non-relativistic approaches\,\cite{PINEDA} but more accurate.




\end{document}